\documentclass{ws-procs961x669}            
\begin{document}
\title{Quantum gravity corrections to the standard model Higgs \\ in Einstein and $R^2$ gravity}

\author{Yugo Abe, Masaatsu Horikoshi}

\address{Department of Physics, Shinshu University, Asahi3-1-1, Matsumoto, Nagano 390-8621, Japan}

\author{Takeo Inami}
\address{Department of Physics, National Taiwan University, Taipei 10617, Taiwan, R.O.C. \\ Theoretical Research Division, Nishina Center, RIKEN, Wako 351-0198, Japan}

\begin{abstract}
 We evaluate quantum gravity corrections to the standard model Higgs potential $V(\phi)$ a la Coleman-Weinberg and examine the stability question of $V(\phi)$ at scales of Planck mass $M_{\rm Pl}$. We compute the gravity one-loop corrections by using the momentum cut-off in Einstein gravity. The gravity corrections affect the potential in a significant manner for the value of $\Lambda= (1 - 3)M_{\rm Pl}.$ In view of reducing the UV cut-off dependence we also make a similar study in the $R^2$ gravity.

\end{abstract}


\bodymatter

\section{Introduction}
~~~It is curious that the mass of the Higgs boson $m_{\rm H}$ ($=125.09\pm0.24\,{\rm GeV}$), which has recently been discovered at LHC, lies far outside of the mass bound derived from the one-loop radiative corrections \cite{CMPP,SZ}.
This bound arises from the stability condition 
on the Higgs quartic coupling $\lambda$, i.e. $\lambda(\mu)>0$.
The large two-loop corrections come into play and the renormalization group (RG) flow of $\lambda$ changes drastically. 
The flow is also tangled with the top quark mass $m_{\rm t}$. 
Some fine-tuning of the parameters, especially that of $m_{\rm t}$, yield the $\lambda$ barely in accord with the boundary values of the stability bound extended to the scales of Planck mass $M_{\rm Pl}$. \cite{HKO,BKKS,DVEEFGA}
It implies an interesting possibility that the standard model may be the correct unification theory all the way up to the Planck scales $M_{\rm Pl}$.\cite{BDGGSSS,BKPV,AFS}
This scenario is consistent with the so far vain results of the susy particle search at the LHC experiment. 

It is a common belief that quantum gravity effects should manifest itself at Planck scales, irrespectively of the correct quantum theory of gravity. Hence it is a very urgent problem to study whether incorporating quantum gravity corrections may change drastically the RG analyses with the matter loop corrections alone.\cite{LP,HKTY,BM} 

In this paper, we calculate the gravity loop corrections to the Higgs potential $V(\phi)$ and examine the property of the Higgs potential at Planck scales $M_{\rm Pl}$. 
In the Einstein gravity, we have found that there is a significant difference between the Higgs potential with and without gravity corrections. 
In the previous works,\cite{LP,BM} the gravity loop corrections to the $\phi^6$ and $\phi^8$ terms are studied.
We consider all such corrections on the RG analysis of $V(\phi)$.
This talk contains a brief summary of our recent paper.\cite{AHI}
\section{Gravitational Coleman-Weinberg corrections}

 In the standard model (SM), one- and two-loop contributions to the bare Higgs mass and the bare couplings have been studied in details. \cite{HKO,BKKS,DVEEFGA}
 We take into account the gravity one-loop effects in addition to the matter loop effects and will see how two contributions compete in the energy region of Planck mass scales $M_{\rm Pl}$.
 
 We derive the Higgs effective potential $V(\phi)$ in the Einstein's gravity theory, by extending the Coleman-Weinberg mechanism \cite{CW} to gravity loops\cite{LP,BM,Smolin}. 
We begin by writing the action for the Higgs field $H$ and the metric $g_{\mu\nu}$
 \begin{equation}
  S=
   \int d^{4}x \sqrt{-g}\left[\frac{2}{\kappa^2}R
   +g^{\mu\nu}(\partial_{\mu}H)^{\dagger}(\partial_{\nu}H)
   +m^{2}H^{\dagger} H-\lambda(H^{\dagger}H)^{2}
   +\cdots \right].
 \end{equation}
 where $\kappa:=\sqrt{32\pi G}=\sqrt{32\pi}M_{\rm Pl}^{-1}$ ($M_{\rm Pl}=1.22\times10^{19} {\rm GeV} $). The ellipses show the terms of gauge and fermion fields.
 Expanding the fields around background fields as $H^{\dagger}=1/\sqrt{2}\, \left(\sigma_1-i\,\pi_1, \phi+\sigma_2-i\,\pi_2\right)$ and $g_{\mu\nu}=\eta_{\mu\nu}+\kappa h_{\mu\nu}$, we evaluate the gravity corrections for the tree level Higgs potential
 \begin{equation}
 \begin{split}
  V_{\rm tree} \ &=\ -\frac{1}{2}m^2\phi^2+\frac{1}{4}\lambda\phi^4\ .
 \end{split}
 \end{equation}
  
 We take the de Donder gauge fixing term $\mathcal{L}_{\rm gf}$. It is given in the Minkowski background by
 \begin{equation}
 \begin{split}
  \mathcal{L}_{\rm gf}\ =\ -\eta_{\alpha\beta}
   \left(\eta^{\mu e}\eta^{\nu \alpha}
   -\frac{1}{2}\eta^{\mu \nu}\eta^{e \alpha}\right)
   \left(\eta^{\rho f}\eta^{\sigma \beta}
   -\frac{1}{2}\eta^{\rho\sigma}\eta^{f \beta}\right)
   h_{\mu\nu ,e}h_{\rho\sigma , f}.
 \end{split}
 \end{equation} 
 
 The Higgs and graviton 1-loop corrections to $V_{\rm tree} (\phi)$ are obtained in the cut-off method. \cite{LP,BM,Smolin}
 \begin{equation}
 \begin{split}
  \delta V_{\rm eff}\ =&\ 
   \frac{3}{64\pi^2}\left(-m^2+\lambda \phi^2 \right)^2\ 
    \left(\ln\frac{-m^2+\lambda\phi^2}{\Lambda^2}-\frac{3}{2}\right)
    \\ &\ 
   +\frac{9}{256\pi^2}\kappa^4\left(-\frac{1}{2}m^2\phi^2
    +\frac{1}{4}\lambda\phi^4\right)^2
    \left(\ln{\left(\frac{\kappa^2}{2}
     \frac{-\frac{1}{2}m^2\phi^2
     +\frac{1}{4}\lambda{\phi}^4}{\Lambda^2}\right)}
     -\frac{3}{2}\right)
    \\ &\ 
   +\sum_{i=\pm}\frac{C_{i}^2}{64\pi^2}\left(\ln\frac{C_{i}}{\Lambda^2}
    -\frac{3}{2}\right) ,
  \label{eq:potential_eff}
 \end{split}
 \end{equation}
 where $C_{\pm}$ is
 \begin{equation}
 \begin{split}
  C_{\pm}\ =\ \frac{1}{2}
  \left[
   m^2_{C}-m^2_{A}\ \pm\sqrt{(m^2_{C}+m^2_{A})^2-16m^4_{B}}\,
  \right] .
 \end{split}
 \end{equation}
 Here
 $m^2_{A}=\frac{\kappa^2}{2} (-\frac{m^2}{2}\phi^2+\frac{\lambda}{4}\phi^4)$, 
 $m^2_{B}=\frac{\kappa}{2}\left(-m^2\phi+\lambda\phi^3\right)$ and 
 $m^2_{C}=-m^2+3\lambda\phi^2$.

 The second and third terms in (\ref{eq:potential_eff}) are due to the graviton loops, which are suppressed by Planck mass at electro-weak scales.
 Gravity corrections give rise to logarithmic divergent terms of $\phi^6$ and $\phi^8$. 
 Such higher power terms are not significant at usual energies, but they may become significant at high energy scales, i.e. $M_{\rm Pl}$.

  The $\phi^2$ and $\phi^4$ terms may be renormalized in the usual way.
 Expanding the third term $\sum_{i=\pm}{C^2_{i}}=m^4_{C}+m^4_{A}-8m^4_{B}$, the counter terms (CT) are found as 
 \begin{equation}
 \begin{split}
  \delta V_{\rm CT}\ =&\ 
   \frac{3}{64\pi^2}\left(-m^2+\lambda \phi^2 \right)^2\ 
    \ln\left(\frac{\Lambda^2}{\mu^2}\right)
   +\frac{1}{64\pi^2}\left(-m^2+3\lambda\phi^2\right)^2\ 
    \ln\left(\frac{\Lambda^2}{\mu^2}\right)
   \\&\ 
   -\frac{\kappa^2}{32\pi^2}\left( m^4\phi^2-2\lambda m^2\phi^4 \right)\ 
    \ln\left(\frac{\Lambda^2}{\mu^2}\right)
   +\frac{5\kappa^4}{512\pi^2}m^4\phi^4\ 
    \ln\left(\frac{\Lambda^2}{\mu^2}\right) .
 \label{eq:counters}
 \end{split}
 \end{equation} 
 We add these gravitational 1-loop corrections to the tree level potential 
 \begin{equation}
 \begin{split}
  V_{\rm eff}\ =\ V_{\rm tree}+\delta V_{\rm eff} +\delta V_{\rm CT}\ .
 \end{split}
 \end{equation}

 \begin{figure}[ht]
   \includegraphics[width=120mm]{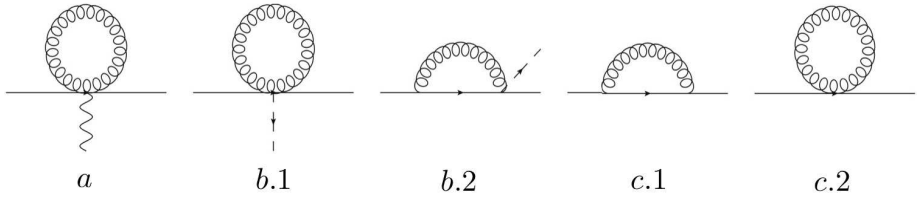} 
    \caption{Gravitational 1-loop diagrams for gauge couplings (a), Yukawa coupling (b) and anomalous dimension of fermion (c).}
    \label{fig:1-loopdiagrams}
 \end{figure}

  The Higgs potential including the loop corrections can easily be obtained by use of RG.
 We already know the $\beta$ functions in SM to the 2-loop order. \cite{HKO,FJJ}
 
 Some graviton loop corrections have recently been computed. \cite{LP,HKTY,BM}
 We have further calculated gravity corrections to other coupling constants, i.e. gauge and Yukawa couplings, as shown in Fig\ref{fig:1-loopdiagrams}.
 $\beta$ functions and anomalous dimensions due to the gravity corrections are
  \begin{equation}
  \begin{split}
   \beta_{m^2 g} & =\ 
    \frac{5\kappa^2 m^2}{16\pi^2}\, \mu^{2}
     -\frac{\kappa^2m^4}{8\pi^2}, \ 
   \beta_{\lambda g} =\ 
    \frac{5\kappa^{2}\lambda}{16\pi^2}\mu^{2}
     -\frac{\kappa^2\lambda m^2}{2\pi^2}
     -\frac{5\kappa^{4}m^{4}}{64\pi^2}, \\
   \beta_{y_t g}\ &=\ 
    \frac{\kappa^2}{2\pi^2}y_t\mu^2 ,\ 
   \gamma_{\phi g}\ =\
    -\frac{\kappa^2 m^2}{32\pi^2} ,\ 
   \gamma_{t g}\ =\
    \frac{27\kappa^2}{512\pi^2}\mu^2. 
   \label{eq:betafunctions} 
  \end{split}
  \end{equation}

\section{The potential due to gravity corrections}

 We solve the RG equations for the coupling constants and $V_{\rm eff}$, using the $\beta$ functions and anomalous dimensions due to SM \cite{HKO} and gravity corrections given in (\ref{eq:betafunctions}). 
 We employ the threshold values of these quantities given in the literature \cite{DVEEFGA}
 \begin{equation}
 \begin{split}
  g_y(m_t)\ &=\ 0.45187
    ,\ 
  g_2(m_t)\ =\ 0.65354
   ,\\
  g_3(m_t)\ &=\ 1.1645-0.00046\left(\frac{m_t -173.15}{\rm GeV}\right)
    ,\\
  y_t(m_t)\ &=\ 0.93587+0.00557\left(\frac{m_t -173.15}{\rm GeV}\right)
   -0.00003\left(\frac{m_H-125}{\rm GeV}\right)
   ,\\
  \lambda (m_t)\ &=\ 0.12577 + 0.00205\left(\frac{m_H-125}{\rm GeV}\right)
   -0.00004\left(\frac{m_t-173.15}{\rm GeV}\right)
   .
 \end{split}
 \end{equation}
 We adjust the value of $m^2(m_t)$ so that $V_{\rm eff}(\phi)$ gives correct the vacuum, $v=246{\rm GeV}$.
  
 \begin{figure}[h]
   \includegraphics[width=60mm]{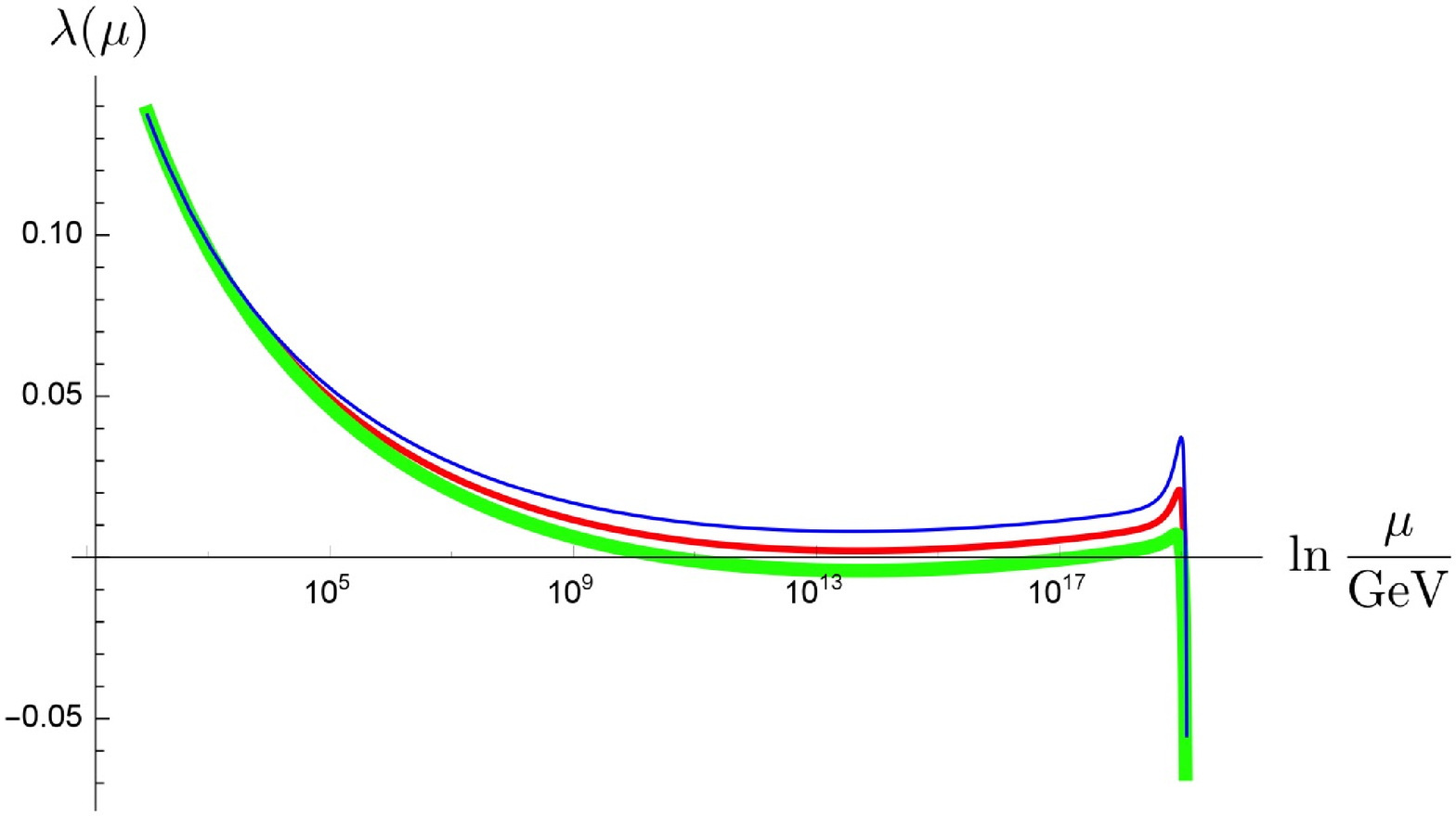}
   \includegraphics[width=60mm]{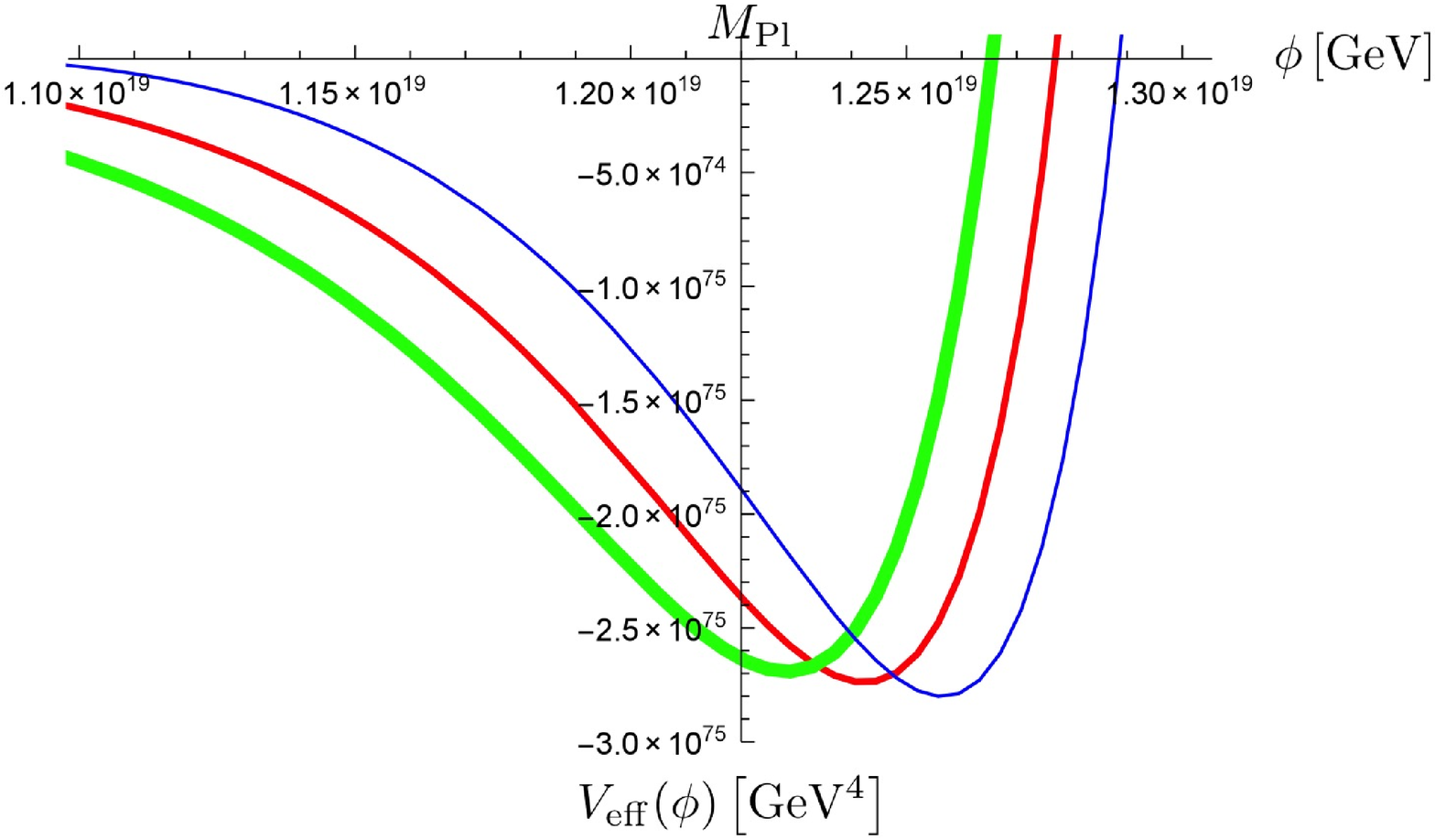}
    \caption{RG flow of $\lambda(\mu)$ and $V_{\rm eff}(\phi)$ at $\phi \sim M_{\rm Pl}$ These curves correspond to top mass values $m_t=174,173,172{\rm GeV}$ in thick and color order, Green, Red, Blue.}
    \label{fig:gcwefftop}
 \end{figure}

 The influence of the gravitational corrections to $\lambda(\mu)$ and $V(\phi)$ is very small, the $\lambda(\mu)$ and $V(\phi)$ are consistent with those in SM, at $\mu , \phi < \mathcal(O) (10^{17}) {\rm GeV}$. 
 
 \vspace{0.2cm}
 \leftline{i) $\mu$-dependence of $\lambda$}
 
 Gravity corrections are noticeable around $\mu =\mathcal{O}(10^{18} {\rm GeV})$, having the effect of increasing the value of $\lambda$ rapidly, as seen in the Fig\ref{fig:gcwefftop}. This behavior stops at $\mu = (0.9 \sim 1.0) \times M_{\rm Pl}$, and they start to decrease $\lambda $ sharply. $\lambda$ becomes negative at $\mu \sim M_{\rm Pl}$.
 
 \vspace{0.2cm}
 \leftline{ii) $V(\phi)$ near $\phi\sim M_{\rm Pl}$}
 
 Regarding $V(\phi)$, gravitational effects begin to be noticeable at $\phi=\mathcal{O}(10^{18} {\rm GeV})$, $\phi^6$ and $\phi^8$ terms become dominant.
 $\phi<M_{\rm Pl}$, $V(\phi)$ is increasing and is positive. 
 $\phi \sim (0.8\sim 0.9)\times M_{\rm Pl}$, $V(\phi)$ begins to fall.
 It stops falling at $\phi=1.1 \times M_{\rm Pl}$, and takes the minimum there.
 At $\phi\sim 10^{20}{\rm GeV}$ and larger, $V(\phi)$ is rapidly increasing.
 However, at such large values of $\phi$, more higher loop effects will be dominant, one cannot say anything reliable about graviton loop corrections.

 \begin{figure}[ht]
  \begin{minipage}{.5\textwidth}
   \includegraphics[width=60mm]{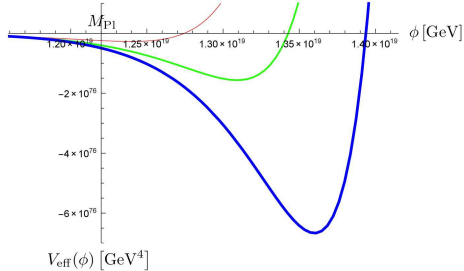}
  \end{minipage}
  \begin{minipage}{.4\textwidth}
    \caption{$V_{\rm eff}(\phi)$ at $\phi \sim M_{\rm Pl}$, $m_{H}=125.09{\rm GeV}$, $m_{t}=173.21{\rm GeV}$ and $\Lambda=3M_{\rm Pl},2M_{\rm Pl},M_{\rm Pl}$ in thick order. The vertical axis is on the position of $\phi=M_{\rm Pl}$.}
    \label{fig:gcweffcut}
  \end{minipage}
 \end{figure}

  \leftline{iii) $\Lambda$ dependence of $V(\phi)$}
  
 At $\phi \sim M_{\rm Pl}$ or larger, $\phi^6$and $\phi^8$ terms are significant but they depend on the cut-off value $\Lambda$, as shown in Fig \ref{fig:gcweffcut}.
 $V(\phi)$ takes the minimum for each cut-off $\Lambda = 1 \times M_{\rm Pl},2 \times M_{\rm Pl}, 3 \times M_{\rm Pl}$.
 One may safely say that $V_{\rm eff}(\phi)$ takes a minimum; the depth of the minimum depends strongly on the cut-off value $\Lambda$, however.
 The value of $\phi$ at the minimum ($\phi_m$) increases proportion ally to cut-off $\Lambda$.
 However $\Lambda$ dependence of $\phi_m$ is mild, $\phi_m$ stays around $M_{\rm Pl}$. 
 Hence, $V (\phi)$ takes the minimum at $\phi < \Lambda$ except the case of  $\Lambda=1\times M_{\rm Pl}$.


\section{Gravity corrections in $R^2$ gravity}
 We have evaluated the quantum gravity corrections to Higgs potential $V(\phi)$ in the Einstein gravity in the cut-off method. 
 The resulting $V(\phi)$ apparently depends on the cut-off $\Lambda$.
 It is desirable to study the gravity corrections in some UV renormalizable modified gravity, so that they have no $\Lambda$ dependence or at most mild $\Lambda$ dependence. 
 One modest approach is to take $R^2$ gravity and evaluate $R^2$ gravity C-W corrections to $V(\phi)$. 
 A different approach has been taken in an early work \cite{Smolin}.  
 A certain $R^2$ gravity, i.e. the action of $R^2$ term $+(R_{\mu\nu})^2$ term is known to be UV renormalizable. \cite{Stelle,FT}
 $R^2$ term alone may not be renormalizable but the UV divergence behavior is mild, and this theory tentatively serves our purpose.
 \begin{equation}
  \begin{split}
   S=
   \int d^{4}x \sqrt{-g}\left[\frac{2}{\kappa^2}\left(R + \alpha R^2\right) 
   +\cdots\right].
  \end{split}
 \end{equation}
 
 In this case, graviton's propagator is $i P_{\mu\nu;\rho\sigma}$, where 
 \begin{equation}
  \begin{split}
   P_{\mu\nu;\rho\sigma}\ =\ 
    \frac{\eta^{\mu\nu}\eta^{\rho\sigma}-\eta^{\mu\rho}\eta^{\nu\sigma}
	-\eta^{\mu\sigma}\eta^{\nu\rho}}{2k^2}
	-\frac{1}{4}\frac{\eta^{\mu\nu}\eta^{\rho\sigma}}{k^2+\frac{1}{4\alpha}}.
  \label{eq:propagator_of_R2}
  \end{split}
 \end{equation}
 
 The difference between the propagator of the $R^2$ gravity and the Einstein gravity is the presence or absence of the second term at (\ref{eq:propagator_of_R2}). 
 Due to the second term of the propagator, we get the UV finite 1-loop effects about $\phi ^6$ term 
 \begin{equation}
  \begin{split}
   -\frac{\kappa^2\lambda^2}{20\pi^4}\ln{\left(4\alpha\mu^2\right)}\,\phi^6\ .
  \end{split}
 \end{equation}
 As expected, the cut-off $\Lambda$ does not appear. The same remark may not hold true for $\phi^8$ and higher terms.  
 

\section{Conclusions}
Evaluating the quantum gravity corrections to the Higgs potential $V(\phi)$ in the Einstein gravity in the cut-off method, we have found a salient difference between $V (\phi)$ with and without gravity corrections in the region $\phi \gtrsim M_{\rm Pl}$. 
$V (\phi)$ with gravity corrections takes a minimum at $\phi \sim M_{\rm Pl}$, while $V(\phi)$ without gravity corrections monotonically increases. 
 $V(\phi)$ depends on the cut-off scales, because the Einstein gravity is not renormalizable.
As seen in Fig \ref{fig:gcweffcut}, $\Lambda$ dependence of the depth of the minimum of $V (\phi)$ is strong.
Whereas, the value of $\phi$ at the minimum depends weakly on the value of $\Lambda$, the minimum exists regardless of the cut-off scales.
 We may safely guess that qualitatively due to gravity corrections, the Higgs potential takes the minimum around the $\phi\sim M_{\rm Pl}$.
 
There is a possibility of that Higgs potential does not depend on the cut-off scales if we consider some modified theories of gravity. 
  In a simple case of $R^2$ gravity ( without $(R_{\mu\nu})^2$ term), the $\phi^6$ term correction does not depend on $\Lambda$.

\section*{Acknowledgment}
 This talk(M.\,H.) is supported in part by funding from Shinshu University.


\end{document}